# Single-shot framing integration photography with high spatial resolution at $5.3\times10^{12}$ frames per second by an inversed 4f system

QIFAN ZHU,[1,2] YI CAI,[1,2] XUANKE ZENG,[1,2] HU LONG,[1,2] LIANGWEI ZENG,[1,2] YONGLE ZHU,[1,2] XIAOWEI LU [1,2,3] AND JINGZHEN LI [1,2,4]

[1] *College of Physics and Optoelectronic Engineering, Shenzhen University, Shenzhen 518060, China*
[2] *Shenzhen Key Laboratory of Micro-Nano Photonic Information Technology, College of Physics and Optoelectronic Engineering, Shenzhen University, Shenzhen 518060, China*
[3] *xiaoweilu@szu.edu.cn*
[4] *lijz@szu.edu.cn*

**Abstract:** We present a framing integration ultrafast photography (FIP), which integrates the framing structure, codes a dynamic event by an inversed 4f system (I4F) and decodes regionally with high spatial resolution. In the experiment about laser-induced plasma, FIP achieved a framing rate of $5.3\times10^{12}$ frames per second (fps) and an intrinsic spatial resolution of 110.4 lp/mm. It has an excellent spatio-temporal resolution and a compact and flexible structure. Hence, it can probe unrepeatable ultrafast intra- and inter-atomic/molecular dynamics, different in size and duration with high-quality. Besides, FIP can lay a foundation for integrating and simplifying ultrafast photography instruments. Here the minimum framing time (temporal resolution) is limited by only the laser pulse duration; be sure, attosecond laser technology may further increase framing rates by several orders of magnitude.

## 1. Introduction

Ultrafast imaging is widely used to record dynamic events at a femtosecond time scale and helps understand fundamental scientific questions, such as biology[1], physics[2], chemistry[3], and so on. Pump-probe methods based on repetitively measured can record repeatable events effectively[4, 5]. However, it cannot record unrepeatable dynamics, such as the process of laser processing[6], detonating, high voltage discharge[7], and so on. Hence, it is meaningful for researches on single-shot ultra-fast imaging methods.

Framing structure plays an essential role in single-shot ultrafast photography. Conventionally, the framing structures are produced by multiple optical paths with different delays, such as sequential holographic imaging (SSSHI)[8, 9], frequency recognition algorithm for multiple exposures (FRAME)[10], noncollinear optical parametric amplification for single-shot imaging (FINCOPA)[11], and so on. These framing structures limit the number of frames, make their system complex and massive, and increase the difficulties for adjustment. In 2018, Ki-Yong Kim developed a single shot ultra-fast imaging at picosecond time scales by mirror array with different angles and delays, integrating the framing structure[12]. However, frames in this method are with different rotations, and the size of mirrors limits fields of view. In 2014, K. Nakagawa provided a sequentially timed all-optical mapping photography (STAMP)[13]. STAMP corresponds the spectrum to the times based on a chirp pulse to produce sequential pulses and realized high resolution and sub-picosecond time scales with the further study[14, 15]. However, STAMP methods lost spectral information and limited the framing time because of Heisenberg's uncertainty principle. In 2019, Chen Hongwei provided a method with a silicon photonic integrated chip, which has the function of multi-angle illumination and provides ultra-short delays for each illumination source[16]. However, the method has process requirements for the light source, and divergence angles are fixed.

Besides, spatial resolution is a significant parameter in single-shot ultrafast photography. It not only suffers from the quality of the optical elements but also some other factors. In recent years, compressed sensing technology is widely used in ultra-fast imaging. Compressed ultrafast photography (CUP) can significantly enhance the number of frames in a single shot[17-20]. However, CUP has a low spatial resolution. Compared with CUP methods, the method based on spatial angular multiplexing and recognition in a frequency domain has the advantages of accurate reproduction. The single-shot sequential holographic imaging method (SSSHI) usually used this method[21, 22]. However, the fields of SSSHI limited suffered from the walk-off affection. There are several methods to reduce the conflict in SSSHI[23, 24], the problem still exists. In 2017, FRAME used structural illumination to solve the problem[10]. Only the pulse width limits the temporal resolution, which can achieve a spatial resolution of ~15 lp/mm but the method with a considerable split structure. A digital mirror Device simplified the vast structure in the following studying, but the walk-off affection problem appeared again, and the temporal resolution was at microsecond time scales[25, 26]. However, the spatial resolution of these methods is limited by the stripe patterns resolution and the number of frames. In 2021, FINCOPA has outstanding spatial and temporal resolutions, which can achieve a spatial resolution of ~83 lp/mm with a magnification of 3× and an effective frame rate of 15 Tfps[11]. Nevertheless, its spatial resolution is limited by noncollinear optical parametric amplification.

This article presents an integration ultrafast photography system based on an inversed 4f system (FIP). FIP has a compact and flexible system, which integrates the framing structure with an inversed 4f system. Its spatial resolution only suffers from the optical elements in this system. Its framing time can be close to a pulse width of an ultrashort pulse. Hence, FIP has excellent performances in the temporal and spatial resolutions.

## 2. Principle and methods

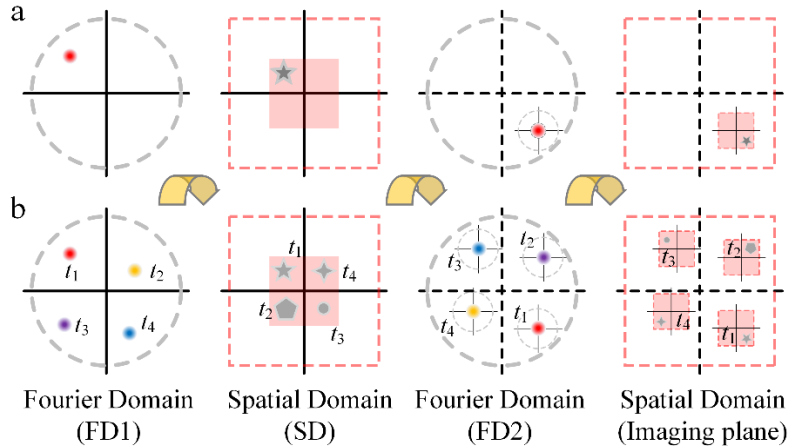

Fig.1 The operating principle of the FIP system: a) the principle of I4F; b) the principle of FIP.

The methodology of inversed 4f system (I4F) is shown in Fig.1a. 4f system, an optical information processing system consisting of two lenses, modulates spatial information in a frequency domain. I4F has the same structure as a 4f system, but it modulates a spatial domain. It can separate the spatial information in the same region into different frequency domain positions with small sizes. Hence, it provides a possibility to integrate the framing structure. A point in a frequency domain (FD1) is transformed into a spatial domain (SD) and illuminates a region in the center of SD. If a point in FD1 illuminates an object, it will be a point with the information of the object at the symmetry position with itself in a frequency domain (FD2). To achieve the object information, we only need to decode the region in the center of the point in FD2. FIP methodology is based on I4F and is shown in Fig.1b. If there are several points at different positions in FD1 with different appearing times, they will illuminate a dynamic event in SD at different moments. Then, there are several symmetry points in FD2 with different times information. Finally, we can decode different times information in different positions.

The FIP system is shown in fig.2a and consisted of three parts: integrated framing structure, I4F system, and imaging structure. The optical elements in the system satisfy the positional relationship shown in fig.2b.

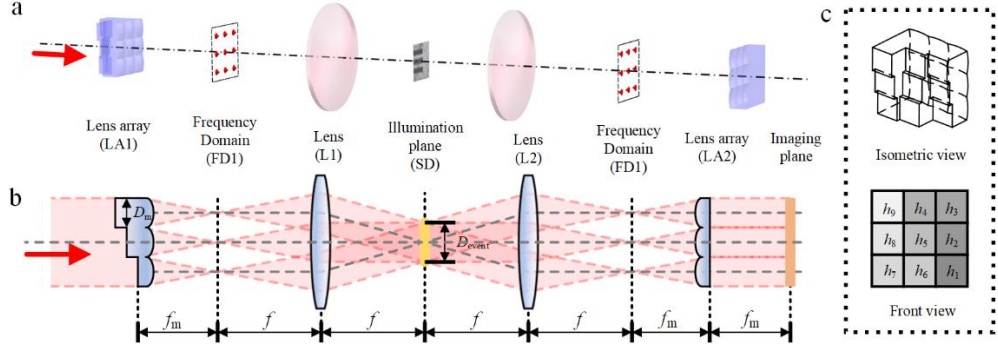

Fig.2 a) The 3D model of FIP system; b) light path diagram of FIP system; c) isometric and front view of framing structure.

The integrated framing structure is composed of a step array and a lens array, shown in Fig.2b. The step array is aimed to realize framing, and the lens array is used to establish a frequency plane for the I4F system. Steps in the step array have different positions and thicknesses. The number of steps decides the number of frames. The thickness differences $\Delta h$ between adjacent steps are the same and determine a framing time

$$t_f = \Delta h(n_g - 1)/c, \qquad (1)$$

$n_g$ is the group index of steps; $c$ is the velocity of light. The step array integrates the framing structure into a single optical element. The lens array (LA1) is applied to establish a frequency plane for FD1. Lenses in LA1 have the same size and focal length. The FD1 is at the back focal plane of LA1, and a plane wave that passes through the LA1 can set up several points in FD1. Because the lenses correspond to the steps one by one, each point in FD1 can appear differently. Then we can establish a time-varying optical field in FD1.

I4F system is made up of two lens L1 and L2. It makes the probe pulses carry the dynamic information at different moments. The front focus of L2 and the back focus of L1 are coincided and are the position of SD. FD1 is at the front focus of L1, FD2 is at the back focus of L2. All the probe pulses illuminate in the same region at the center in SD; the shape of the illuminated region is decided by the shape of sub-lens, whose sizes are satisfied

$$D_{SDx} = \frac{f}{f_m} D_x, D_{SDy} = \frac{f}{f_m} D_y. \qquad (2)$$

In the illuminated region, the illuminated dynamic event modulates the probe pulses. Then, we can get a new optical field in FD2, which retains the related position relationship in FD1 and carries information of the dynamic event at different times. These dynamic event's information at different times are shifted with each probe pulse in FD2.

The imaging structure is made up of a lens array (LA2) and a detector. In FD2, different time information is at different positions. Suppose we use a lens only to transform a point in FD2. In that case, we can get the image of the dynamic event at the corresponding moment. The region for imaging is the same as the sizes of sub-lens in LA1, the image of the dynamic event is inverted, and the size is magnified $f_m/f$ times.

Hence, we adopt the same lens array LA2 as LA1, whose front focal point is coincidental with the back focal point of L2. LA2 can image all the information with different times in a single-shot at the back focal. Because the information is decoded by optical elements directly, we can get high-spatial-resolution images.

## 3. Simulation and Experiment

### 3.1 Simulation

The simulation structure is shown in Fig.2a and b. There are nine lenses in LA1 and LA2, and we can get nine frames in a single-shot. These lenses have sizes of 3×3 mm and a focal length of 200 mm. The thicknesses of the steps in the step array distribute as Fig.2c. L1 and L2 in I4F are with a focal length of 200 mm. The dynamic event is shown in Fig.3a, which is consisted of nine frames in SD and added random phase distributions shown in Fig.3b. The incident light is a plane wave whose wavelength is at 800 nm, and pulse width is very short and normally passes through the step array. We assume that framing time produced by the steps is much longer than the pulse width and ignore affection on pulse width from steps. Hence, we simulate the pulses from lenses in LA1 successively for the dynamic event and superpose them.

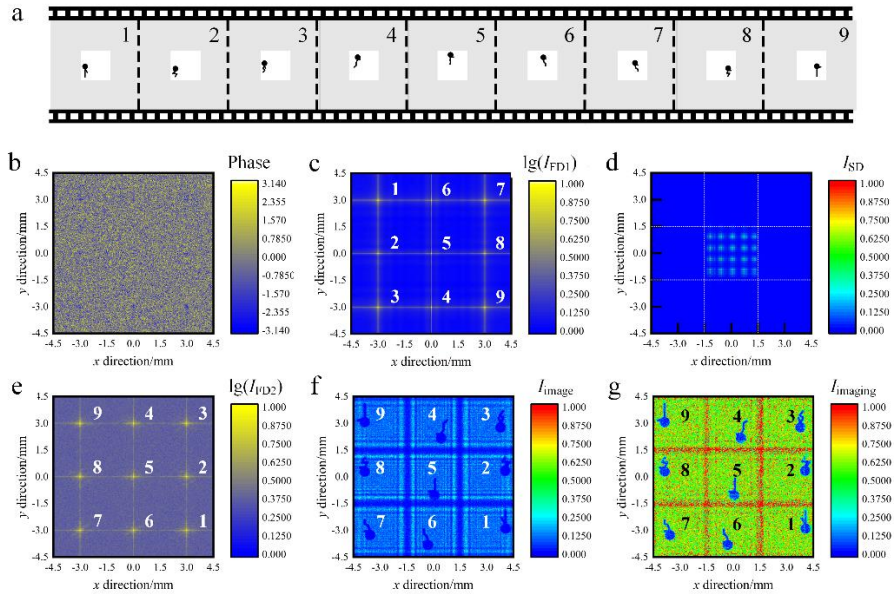

Fig.3 The system was simulated in a single shot, a) the intensity distribution in FD1; b) the intensity distribution in SD; c) the intensity distribution in FD2; d) the phase distribution in FD2; e) the intensity distribution in imaging plane; f) the related intensity distribution after processing; g) 9 frames of the simulated dynamic event.

We simulate a FIP system based on scalar diffraction theory by software MATLAB. The optical field distributions $I_{FD1}$, $I_{SD}$, $I_{FD2}$, $I_{image}$ in FD1, SD, FD2, and imaging plane were achieved. In order to facilitate studying, $I_{FD1}$ and $I_{FD2}$ are shown by an 'lg' function to stick out the patterns. We can find there are nine bright points in the center positions of sub-lenses in LA1 (Fig.3b). In the SD plane (Fig.3d), the optical field distribution is concentrated on the center; sizes of the region are the same as the lens' in LA1, which accords to equ.2. After illuminating a dynamic event, the optical field distribution became nine points with the event information corresponding to the center of lenses in LA1 (Fig.3e). LA2 decodes each point in FD2 and image the frames in the imaging plane (Fig.3f). The images in the imaging plane are inversed. Because of apertures, there are some affections of diffraction for imaging. To reducing the affections, we get frames without events $I_0$ and process the image by the following function

$$I_{\text{imaging}} = \frac{I_{\text{image}}}{I_0}. \tag{3}$$

Then, we get clear and contrasted frames, shown in Fig.3g. The images are the same as Fig.3a and are inversed. Hence, this simulation verifies the feasibility of the FIP system in theory.

*3.2 Experiment*

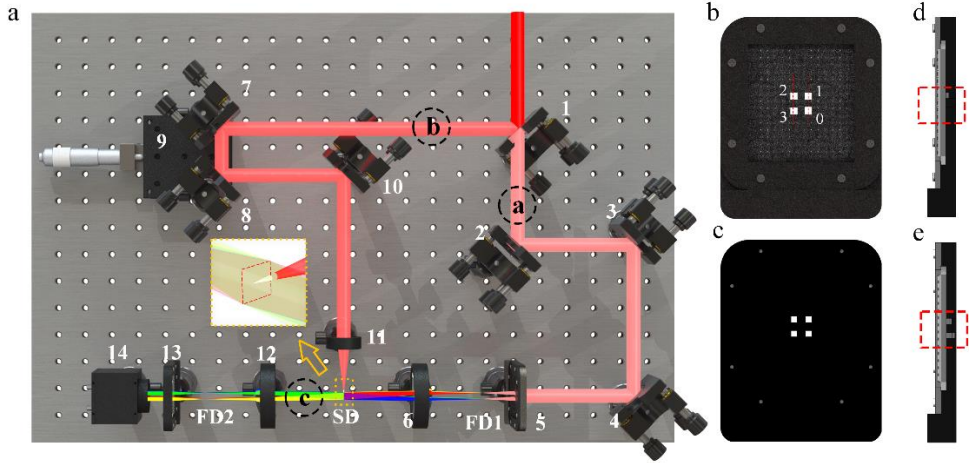

Fig.4 a) Schematic diagram of experimental set-up: (1). beam splitter mirror; elements (2), (3), (4), (7), (8), (10) are mirrors; (5). Integrated framing structure: LA1 and step array; (6). L1; (9). translation stage; (11). convergent lens; (12). L2; (13). LA2; (14). detector. b) left view of the integrated framing structure; c) right view of the integrated framing structure; d) section view at steps 1 and 2; e) section view at steps 3 and 4.

FIP system experiment optical path is shown in Fig.4a. It consists of three parts: (a) probe optical path, (b) excitation optical path, and (c) imaging optical path. The incident light is femtosecond pulse with a wavelength of 800 nm, a pulse width of 35 fs, a spot's diameter of 12 mm, and a repetition frequency of 1KHz. The element (1) divides the incident pulse into probe optical path and excitation optical path.

The probe optical path is made up of elements (2), (3), (4), (5), and (6). Elements (2), (3), and (4) are mirrors and are used for adjusting the optical path. Element (5) is the integrating framing structure, shown in Fig.4b, c, d, and e. LA1 is a lens array with a focal length of 24.5 mm and 4×2.5 mm sizes for each lens (Fig.4b). We instead step array by glass sheets for simplified experimental structure and selected four lenses in LA1 with four frames by 4 square apertures in the LA1 pedestal (Fig.4c). The square apertures and glass sheets have the same sizes of 2×2 mm in the center of each selected lens. Glass sheets are put in these apertures next to LA1, and the numbers in each aperture are shown in Fig.4c. In Fig.4d and e, we can find that the glass sheets in each aperture provide delay times for the framing structure. Each glass sheet provides a delay time of 187 fs with a thickness of 0.12 mm, and material of $SiO_2$ whose group index of 1.4671 at wavelength 800nm. Element (6), L1, is a lens for focus the pulses from LA1 and illuminate in SD. L1 has a focal length of 60 mm, whose back and front focal points coincide with the front focal points of LA1 and SD planes, respectively.

The excitation optical path consists of elements (7), (8), (9), (10), and (11). Elements (7), (8), and (10) are mirrors. Element (9) is a precise translation stage. Element (7), (8), and (9) make up a delay structure which makes the optical length of the excitation optical path equal to probe optical paths. Element (11) is a lens with a focal length of 30 mm and excites a dynamic event about laser-induced plasma in SD.

The optical imaging path consists of elements (12), (13), (14). Element (12) is L2, the same as L1, whose back focal point coincides with the SD plane. Element (13) is LA2. LA2 has the same parameters as LA1, whose back focal point coincides with the front focal point of L2 and the front focal point is at a detector. Element (14) is a detector. It has a resolution of 4088×3072 pixels with 3.1×3.1 μm for each pixel. We set the exposure time of 1 ms, equal to the time interval between two pulses from the laser. Hence, we can get a dynamic event in a single-shot.

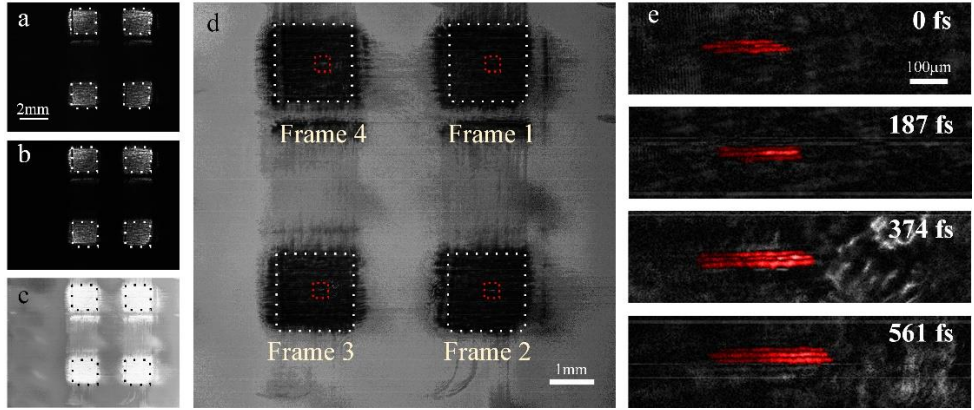

Fig.5 a) four frames with dynamic of laser-induced plasma; b) four frames without dynamic events; c) the image processed based on equ.3; d) the image with further processing; e) dynamic of laser-induced plasma with framing time 187fs.

The experiment results are shown in Fig.5. Fig.5a and b are the images with and without the dynamic of laser-induced plasma in a single-shot, respectively. There are four bright regions in the images, and the size of each region is 2×2 mm, which is equal to the size of the apertures in LA1. Because of the diffraction effects, we cannot observe the dynamic event clearly but can see fuzzy images in each frame. Therefore, we processed the image based on equ.3. However, the four regions are too bright (Fig.5c) to find the dynamic events. The image (Fig.5c) was processed by 1-$I_{imaging}$, shown in Fig.5d. The dynamic of laser-induced plasma is in red rectangle regions. The time order of these regions is shown in Fig. 5d. The four frames dynamic about laser-induced plasma with framing time 187 fs (framing rates $5.3 \times 10^{12}$ fps) are shown in Fig.5e. The four dynamic events were marked red since the laser spot joggling causes some noises, and we can observe an obvious dynamic process. These results verified that the FIP system could achieve dynamic information in a single-shot.

## 4. Discussion

### 4.1 Temporal resolution

Exposure time influences on quality of images. It depends on the pulse width $t_0$, when incident normally to the event. However, pulses almost incident into the event obliquely, shown in Fig.6a. Exposure time suffers from illuminated region $D_{event}$ and illuminated angle $\theta$ ($\theta = f/D_c$), where $D_c$ is the distance between the lens' optical axis in LA1 and the system's, which can be longer than the pulse width. We simulated a dynamic event based on the structure described in section 3.1 without a step array. All nine frames imaged the same dynamic event for further research on these affections. The dynamic is shown in Fig.6b; three squares move from left to right, represented by black squares, and the grey region is the motion trail of objects. This dynamic event is assumed to last one picosecond. A femtosecond pulse, $t_0 = 35$ fs, travels along the z-axis and meets with the center of the dynamic event at 0 fs.

The result of the nine frames is shown in Fig.6c. Red dotted squares are the object at 0 fs, and red arrows indicate the direction when the illumination is projected on the x-y plane. The relationship between the direction of illumination and the object's moving direction affects the quality of frames: 1) Frame 5 is imaged by a coaxial system and has no distortion. It is the clearest that suffers from the pulse width. 2) When the illuminated direction along the motor direction (frame 1,2,3), the frames are the fuzziest in the motor direction (the length of the object in the x-axis). 3) When the illuminated direction along with the minus motor direction (frame 7,8,9), the frames are just fuzzier than frame 5. 4) When the projected illumination direction is perpendicular to the motor direction (frame 4,5,6), the frames appear the object at different times in the perpendicular direction. The exposure time be expanded

$$t_e = t_0 + (t_{front} - t_{back}) = t_0 + \frac{D_{cmax}}{c}\sin\theta, \tag{4}$$

$D_{cmax}$ is the farthest distance between the lens' optical axis in LA1 and the system. Equ.4 indicates that the exposure time mainly suffers from the pulse width and is limited by the illumination angle. If we reduce the illumination angle, the exposure time can be almost equal to the pulse width.

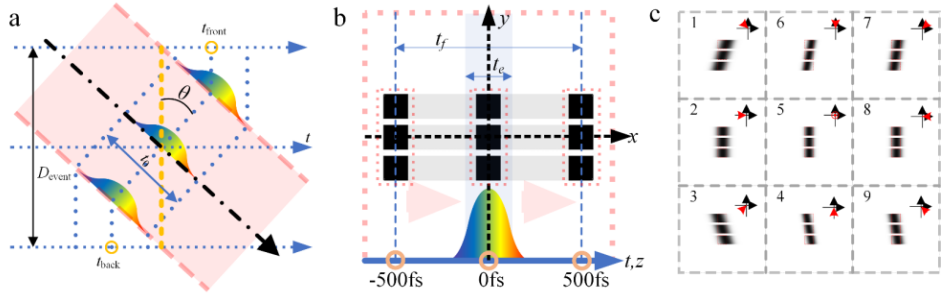

Fig.6 a) Illumination schematic diagram; b) dynamic event schematic diagram. c) the results of the nine frames simulation with different directions of incidence projecting on x-y plane, along: 1) -45°; 2) x-axis; 3) 45; 4) y-axis; 5) z-axis; 6) minus y-axis; 7) -135°; 8) minus x-axis; 9) 135°.

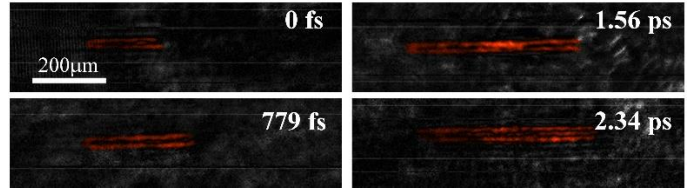

Fig.7 Dynamic of laser-induced plasma with 779 fs framing time

Framing time determines the framing rate and speed of an ultrafast system. Temporal qualify factor $g = t_f/t_e$, which should be equal or greater than 1. The extremity framing rate is limited by exposure. Hence, in a FIP system, minimum framing time should satisfy

$$t_{f\_min} \geq t_e. \tag{5}$$

In a FIP system, minimum framing time $t_{f\,min}$ depends on the thicknesses of steps in the step array. In this experiment, the framing time depends on the thicknesses of glass sheets. Because the minimum framing time can approximate the pulse width, attosecond laser technology may further increase framing rates of FIP by several orders of magnitude.

Furthermore, we experimented with 0.5mm thicknesses of a glass sheet for a laser-induced plasma with a framing time of 779 fs, shown in Fig.7, and the object was also marked red. This result shows that the FIP system can be easy to change the framing time. Hence, FIP suits dynamic events with different time ranges.

## 4.2 Spatial resolution

Spatial resolution is an essential parameter for an imaging system. The imaging part influences the spatial resolution in a FIP system, as shown in Fig.8a. The diaphragm for each frame is the sizes of sub-lenses in LA2. Our experiment measured four frames' spatial resolution by a USAF 1951 optical resolution (Fig.8c and d). The spatial resolutions in imaging plane are 45.25 lp/mm with a field of view (FOV) of 4.9×4.9 mm and magnification 0.41×. The intrinsic spatial resolution of FIP is 110.4 lp/mm. We simulated this real system's MTF (Modulation Transfer Function), shown in Fig.8b; the black line and dotted line indicate the diffraction limit lines for tangential and sagittal planes, respectively. Besides, the other colorful lines and dotted lines are the MTF at different fields. We can find that this system has a very high spatial resolution, but the optical elements reduce the actual spatial resolution. Therefore, FIP can improve the resolution by optimizing the system and increasing the numerical aperture.

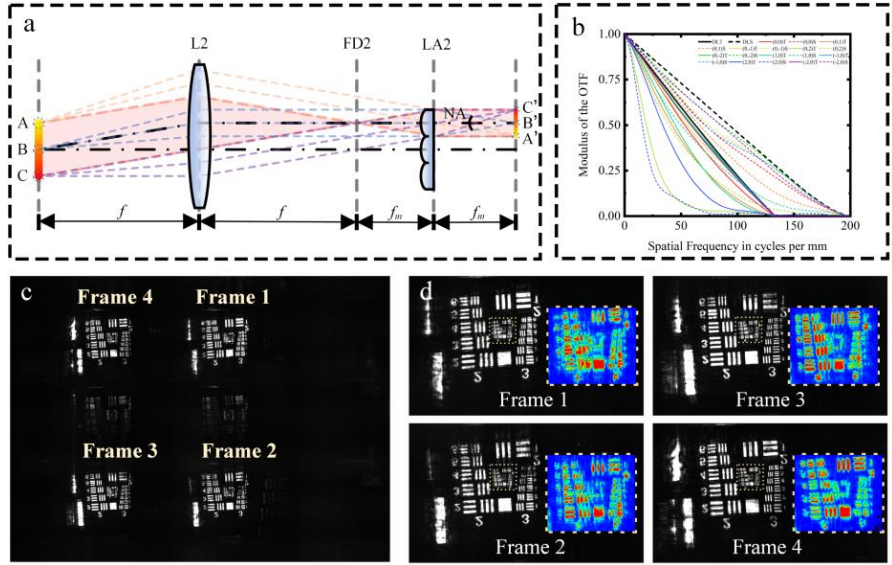

Fig.8 a) Imaging structure schematic diagram; b) MTF of the experiment system; c) the measured image; d) details for four frames (Colorful maps in each frame show the group 4 and 5 of the USAF 1951 optical resolution).

## 4.3 Field of view (FOV)

Field of view (FOV) depends on the size of the illuminated region, according to equ.2. In the laser-induced plasma experiment, FOV is a rectangle region with sizes of 4.9×4.9 mm. We also did a static experiment with only one frame. The sub-lenses of LA1 and LA2 were equivalent to two single lenses with a focal length of 200 mm. These two lenses (L1 and 2) with focus length 400 mm worked as L1 and L2. Incident light is at wavelength 532 nm with a diameter of 8 mm, Fig.9a. In the SD plane, the illuminated spot with a diameter of 16 mm (Fig.9b). We can get an image shown in Fig.9c. Because of the image plane size limiting, the image has a 7.2×5.4 mm size for a FOV of 14.4×10.8mm and magnification 0.5×. FOV of the FIP system can be controlled by shapes and sizes of sub-lens in LA1 and the focal length of L1. Therefore, a FIP system can change FOV based on different objects easily.

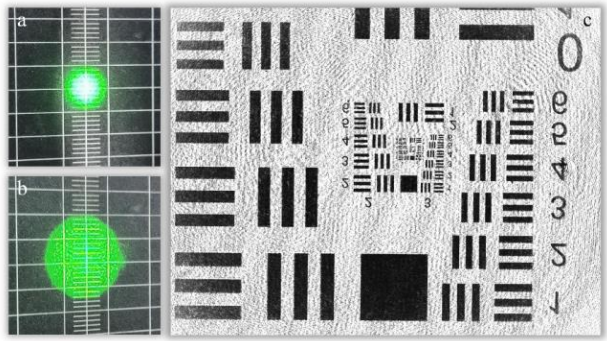

Fig.9 a) incident light spot; b) illuminated light spot; c) image in detector.

### 4.4 Frame number

The number of frames is decided by the number of illuminated lenses in LA1 and the size of the detector. We did an experiment that used the structure described in section 3.2.1 without step array and apertures. There are 12 frames in a single shot (Fig.10a). The right parts of frames 1,2,3,4 are much darker than the left because these frames are at the edge of the incident light spot. Frame 1,5,9,4,8 and 12 are clipped because of the limitation of the detector. If an incident light spot and a detector are large enough, we can get more frames in a single shot.

Besides, we provide a comprehensive system, shown in Fig.10b. When L2 and LA2 adopt different focal lengths with L1 and L2, the frequency distribution in FD2 can remain relationship positions with much larger sizes. Then, we can use multi-detectors to decode points with object information one by one. In this way, we can eliminate the limitation of one detector size and achieve more details for each frame.

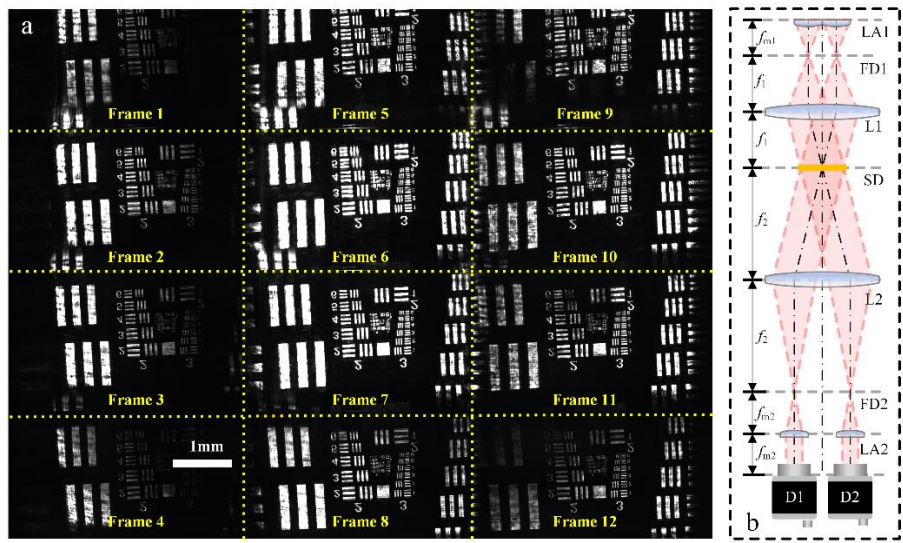

Fig.10 a) Image in detector without apertures in LA1; b) comprehensive system.

## 5. Conclusion

In this paper, we provide a single-shot ultrafast photography system, FIP. It recorded a dynamic event about laser-induced plasma in a single-shot with a framing rate of $5.3 \times 10^{12}$ fps, the intrinsic spatial resolution of 110.4 lp/mm. FIP has a high spatial resolution which only suffers from the optical elements and can be optimized. Because the pulse width mainly limits its framing time, it can be nearly equal to the pulse. Hence, FIP can record dynamic events with high temporal and spatial resolution and be widely used in physics, biology, and chemistry, like the process of laser-induced plasma, propagation of light pulses, atom motion, and so on. If the laser pulse turns into an attosecond time scale, it may further increase framing rates of FIP by several orders of magnitude.

Furthermore, FIP has a compact and flexible structure. It integrates the framing structure by an inversed 4f system, which makes its structure compact. Hence, the elements in FIP will not increase with more frames achieved in a single-shot. Besides, FIP can conveniently adjust system FOV and framing time to suit dynamic events with different sizes and duration times. It also can expand the system to get different magnification images and more details. Hence, FIP lays a foundation for integrating and simplifying ultrafast photography instruments.

## Funding

## Disclosures

The authors declare no conflicts of interest.